\documentclass[epj]{svjour}

\usepackage{graphicx}
\usepackage{fancyhdr}
\def\makeheadbox{{%  remove EPJC header
 }}
\def\mytitle{My title} 
\def\myauthors{My name}  
\def\mytype{My type of session}
\def\mysession{My session}
\def\mytitle{Hidden Fermion in Stueckelberg $Z'$ Models \\as Milli-charged Dark Matter} %Put your title here!
\def\myauthors{Kingman Cheung and Tzu-Chiang Yuan}    %Put your name here!
\def\mytype{Contributed Talk}    
\def\mysession{Alternatives}
\begin{document}
\title{\mytitle}
\author{Kingman Cheung\inst{1,2}
 \and
 Tzu-Chiang Yuan\inst{2}
}                     % Do not remove
\institute{Department of Physics, National Tsing Hua University, Hsinchu 300, Taiwan
\and Physics Division, National Center for Theoretical Sciences, Hsinchu 300, Taiwan}
\date{}
\abstract{
The hidden Stueckelberg $Z'$ model
is augmented by a pair of Dirac fermions. 
If the $Z'$ has a coupling strength comparable to weak scale coupling,
one can show that this hidden
fermion-antifermion pair could be a milli-charged dark matter
candidate with a viable relic density.
Monochromatic photon flux coming from the
Galactic center due to pair annihilation of these milli-charged particles 
is calculated and is shown to be within reach of the next generation
gamma-ray experiments. 
Characteristic collider signature of this theoretical endeavor is that the 
Stueckelberg $Z'$ boson has a large invisible width decaying into the hidden 
fermion-antifermion pair if kinematically allowed.
Singly production of this $Z'$ boson
at the LHC and ILC are studied.
 % end of PACS codes
} %end of abstract
\maketitle
\section{Stueckelberg $Z'$ Extension of Standard Model (StSM)}
\label{intro}
The standard electroweak model (SM) based on the gauge group 
$SU_L(1) \times U_Y(1)$
was extended in \cite{Kors-Nath} by including an additional hidden 
$U_X(1)$ endowed with two Stueckelberg mass terms for the two abelian groups (StSM),
\begin{eqnarray}
{\cal L}_{\rm StSM} &=& {\cal L}_{\rm SM} + {\cal L}_{\rm St} \nonumber\\
{\cal L}_{\rm St}  & = & -\frac{1}{4}C_{\mu\nu}C^{\mu\nu} + \frac{1}{2}(\partial_\mu \sigma + M_1 
C_\mu +   M_2 B_\mu)^2 \nonumber \\
& & - g_X C_\mu {\cal J}^\mu_{X}  \; ,
\end{eqnarray}
where $\sigma$ is the Stueckelberg axion field, $B_\mu$ and $C_\mu$ are the gauge fields for the 
hypercharge $U_Y(1)$ and hidden $U_X(1)$ respectively, and ${\cal J}^\mu_X$ is the current associated with
the hidden gauge field. 
This model is unitary and renormalizable. Upon electroweak
symmetry breaking by the Higgs mechanism 
$\langle \Phi \rangle = v/\sqrt{2}$, the $3 \times 3$  mass matrix for the neutral gauge bosons 
takes the following form
\begin{eqnarray}
\label{massmatrix}
M^2_{\mathrm{St}} & = &\left( \begin{array}{ccc}
  M_1^2 & M_1 M_2 & 0 \\
 M_1 M_2 & M_2^2  + \frac{1}{4} g_Y^2 v^2 & -\frac{1}{4} g_2 g_Y v^2  \\
 0 & -\frac{1}{4} g_2 g_Y v^2  & \frac{1}{4} g_2^2 v^2 \end{array} \right ) 
\end{eqnarray}
where $g_2$ and $g_Y$ are the two gauge couplings for the $SU_L(2)$ 
and the hypercharge  $U_Y(1)$ respectively. Note that the vanishing of the determinant of 
$M^2_{\mathrm{St}}$ guarantees a zero mode that can be identified as the photon. 
For the other two massive states, one is the standard model $Z$ and the other a new Stueckelberg $Z'$. 
Using the precision electroweak data from LEP to constrain the $Z$ mass shift, 
it was shown in \cite{Feldman-Liu-Nath-1} that 
$$
\vert \delta \equiv \tan \phi = M_2 / M_1 \vert \leq 
0.061 \, \sqrt{1- (m_Z/M_1)^2}
$$
assuming  $m_Z / M_1\ll 1$. Furthermore, if one assumes ${\cal J}_X = 0$, the $Z'$ couplings to the SM fermions will be suppressed by the mixing angles which parameterize the orthogonal matrix that diagonalizes the mass matrix (\ref{massmatrix}).
This implies a very narrow $Z'$ width, typically of order $\mathrm{MeV}$. 
Consequently, the Drell-Yan data from the Tevatron can only
constrain the Stueckelberg $Z'$ mass as
\begin{eqnarray}
  m_{Z'} > 250 \;{\rm GeV} \;\; &{\rm for}& \;\; \delta \approx 0.035 \; ,
\nonumber \\
%& \; & \\
  m_{Z'} > 375 \;{\rm GeV} \;\; &{\rm for}& \;\; \delta \approx 0.06 \;,
  \nonumber
\end{eqnarray}
which are much lower than those arising from other more conventional $Z'$ 
models \cite{Feldman-Liu-Nath-1}. 

\section{Hidden Fermions}
\label{milli}

\begin{figure}
\includegraphics[width=0.5\textwidth,height=0.35\textwidth,angle=0]{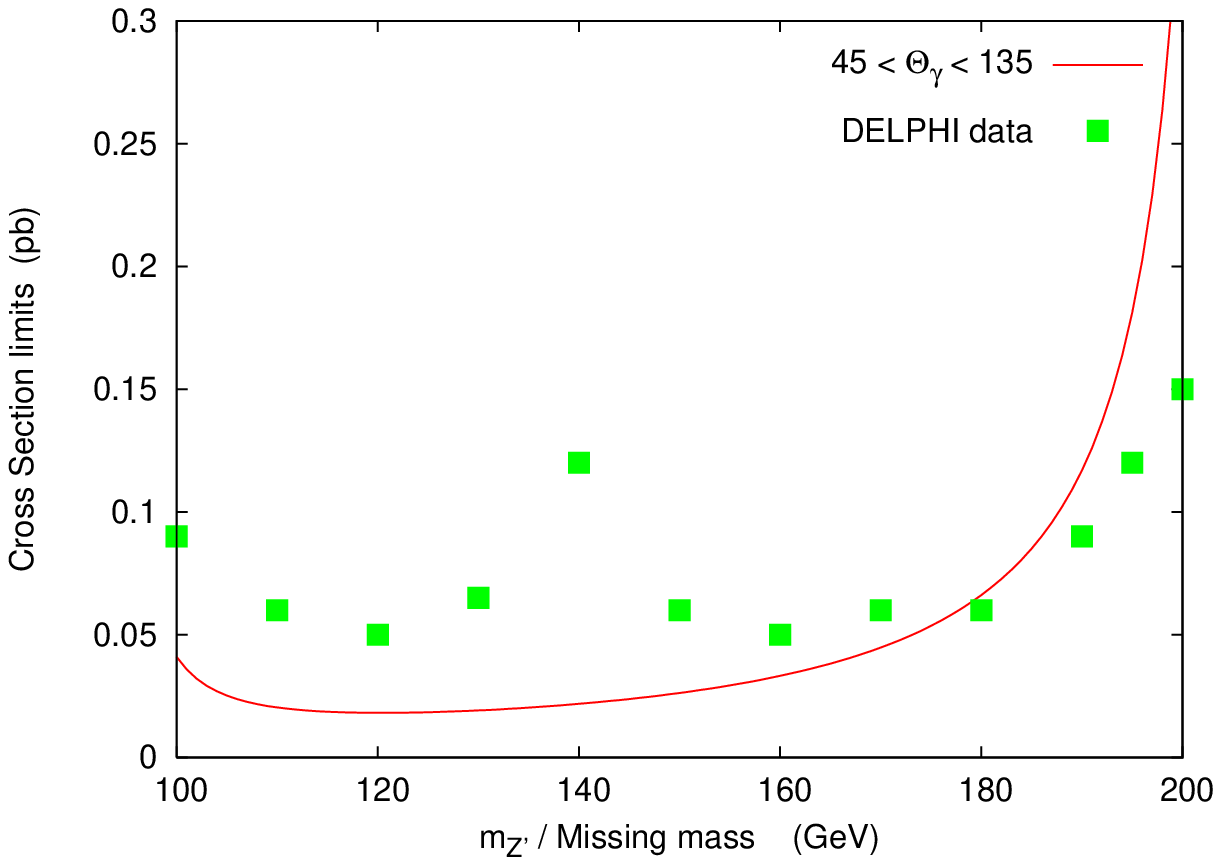}
\caption{LEPII constraint from $e^- e^+ \to Z' \gamma \to \gamma$ + missing energy with 
$m_\chi = 60$ GeV.}
\label{fig:delphi} 
\end{figure}

\begin{figure}
\includegraphics[width=0.5\textwidth,height=0.35\textwidth,angle=0]{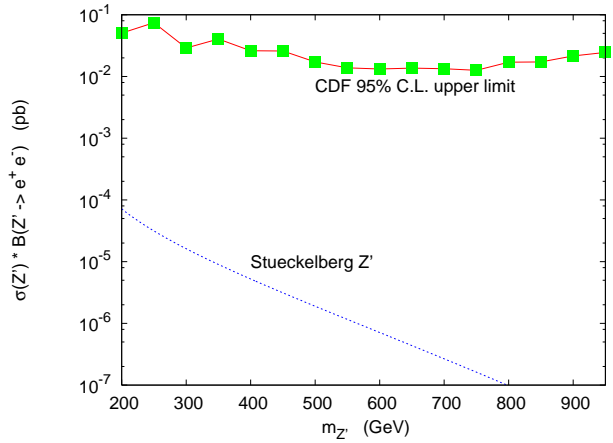}
\caption{Constraint from preliminary CDF data for the Drell-Yan process $p \bar p \to Z' \to e^- e^+$.}
\label{fig:cdf-z} 
\end{figure}

\begin{figure}
\includegraphics[width=0.5\textwidth,height=0.35\textwidth,angle=0]{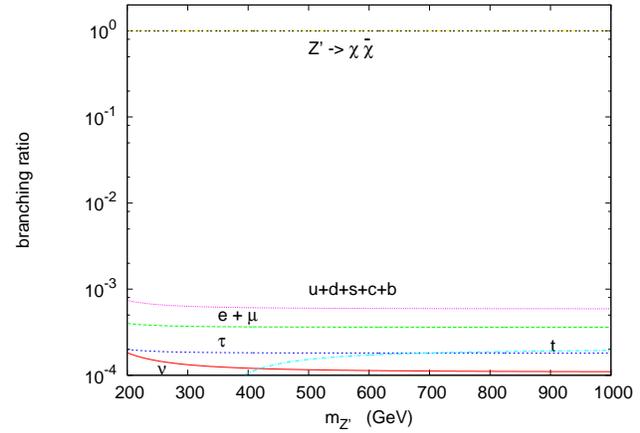}
\caption{Branching ratios for the $Z'$ with $g_X Q^\chi_X = g_2$, $\delta = 0.03$ and $m_\chi = 60$ GeV.}
\label{fig:zprime-br}
\end{figure}

In \cite{Cheung-Yuan}, a hidden Dirac fermion $\chi$ and its antiparticle $\bar \chi$ were 
added in the above StSM such that the hidden current 
${\cal J}_X^\mu = \bar \chi \gamma^\mu Q^\chi_X \chi$ is no longer vanish. After rotation to the 
physical mass eigenstates of the neutral gauge bosons, 
this hidden fermion couples not only to the Stueckelberg hidden $Z'$ 
but also the photon $A$ and the SM $Z$ boson as well. Thus the neutral current becomes
\begin{eqnarray}
\label{neutralcurrent}
- {\cal L}^{NC}_{\rm int} &=& \cdot \cdot \cdot + g_X C_\mu {\cal J}^{\mu\chi}_X \nonumber\\
&=& \cdot \cdot \cdot +
 \bar \chi \gamma^\mu \left[ 
   \epsilon^\chi_\gamma A_\mu 
 +\epsilon^\chi_Z Z_\mu
 +\epsilon^\chi_{Z'} Z'_\mu \right ]\, \chi  
\end{eqnarray}
with the coupling strengths
$  \epsilon^\chi_\gamma = -g_X Q^\chi_{X} c_\theta  s_\phi$,
$  \epsilon^\chi_Z =  g_X Q^\chi_{X} (  s_\psi c_\phi + 
                 s_\theta s_\phi  c_\psi)$ and
$\epsilon^\chi_{Z'} \, = \, g_X Q^\chi_{X} ( c_\psi c_\phi -   s_\theta s_\phi s_\psi)$
where $c_\theta = \cos \theta$, $s_\theta = \sin \theta$, etc are the mixing angles.
If one assumes $g_X Q^\chi_X \sim g_2$, then $\epsilon^\chi_{Z'} \sim g_2$ for small mixing 
angles. This implies the $Z'$ width is no longer narrow if the channel $Z' \to \chi \bar \chi$ is open. 
For typical input values of parameters, the $Z'$ width is of order GeV. 
Experimental constraints on the $Z'$ mass can be further loosened in this scenario. We can 
illustrate this by two plots: Fig.[\ref{fig:delphi}] for 
$e^- e^+ \to Z' \gamma \to \gamma$ + missing energy 
from DELPHI data \cite{delphi} of LEP
and 
Fig.[\ref{fig:cdf-z}] for Drell-Yan process
$p \bar p \to Z' \to e^- e^+$ from the CDF preliminary data \cite{cdf-z} of Tevatron.  
Branching ratio of the $Z'$ is also plotted in Fig.[\ref{fig:zprime-br}] for 
$m_\chi = 60$ GeV.
Once the decay channel $Z' \to \chi \bar \chi$ is open, it dominates over all other modes.

On the other hand, the mixing effects induce an electric charge  $\epsilon^\chi_\gamma / e$
for the hidden fermion $\chi$. This charge is not quantized and can be in the 
interesting milli-charge range depending on the mixing angles. Since $\chi$ is stable in the present model, this provides an alternative milli-charged dark matter candidate for astrophysics.
Milli-charged fermion had been considered as dark matter candidate some time ago \cite{Goldberg-Hall} in a different context \cite{Holdom}.
See also \cite{champs} and \cite{khlopov} for integral and composite charged dark matter 
respectively.
For various experimental limits on milli-charged particles, see \cite{davidson}.

\section{Astrophysical Implication}
\label{astro}

\subsection{Relic density calculation}

\begin{figure}
\includegraphics[width=0.5\textwidth,height=0.35\textwidth,angle=0]{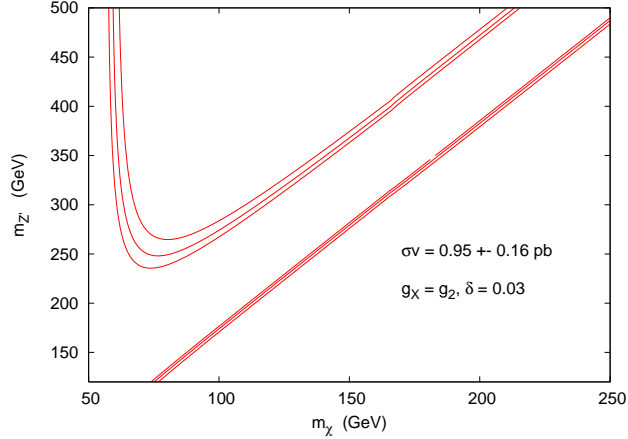}
\caption{Contour plot of $\sigma v = 0.95 \pm 0.08$ pb in the $(m_{Z'}, m_\chi)$ plane 
with $g_X Q_X^\chi = g_2$ and $\delta = 0.03$.}
\label{fig:contourplot} 
\end{figure}

Recent  three year integrated WMAP data \cite{wmap} implies the following constraint for any relic of cold dark matter (CDM)
in our universe
\begin{eqnarray}
\label{wmap}
  \Omega_{\mathrm{CDM}} h^2 & = & 0.1045^{+0.0072}_{-0.0095}  \;\; .
\end{eqnarray}
The relic density is related to the annihilation cross section $\sigma v$ by the following 
approximation \cite{jungman}
\begin{eqnarray}
  \Omega_{\mathrm{CDM}} h^2 
  & \simeq & \frac{0.1 \; {\rm pb}}{\langle \sigma v  \rangle} \; .
\end{eqnarray}
From Eq.(\ref{wmap}), one can translate the WMAP constraint onto the annihilation cross section
\begin{eqnarray}
\label{contour}
 \; \langle \sigma v \rangle \simeq 0.95 \pm 0.08 \; {\rm pb} \; .
\end{eqnarray}
In our relic density calculation \cite{Cheung-Yuan}, we include the following processes 
 $\chi \bar \chi \to f_{\rm SM} {\bar f}_{\rm SM}, \gamma Z'$ and $Z Z'$. 
Thermal average in $\sigma v$ was ignored and $v^2 \simeq 0.1$ was used for the 
relative velocity of the annihilating dark matter.
In Fig.[\ref{fig:contourplot}], the contour plot of $\sigma v$ as restricted by (\ref{contour})
is exhibited as function of the two unknown masses $m_\chi$ and $m_{Z'}$ with 
$g_X Q_X^\chi = g_2$ and $\delta = 0.03$. The upper curves and lower straight lines 
correspond to the case where 
the channel $Z' \to \chi \bar \chi$ is open and close respectively.
One can see that the allowed region for the masses for both $Z'$ and $\chi$ 
are accessible by the LHC. More elaborated calculation of the relic density including proper thermal averaging of $\langle \sigma v \rangle$ over the resonant $Z'$ 
carried out in \cite{Feldman-Liu-Nath-2} do not 
alter our result significantly. 

\subsection{Indirect search}

\begin{figure}
\includegraphics[width=0.5\textwidth,height=0.35\textwidth,angle=0]{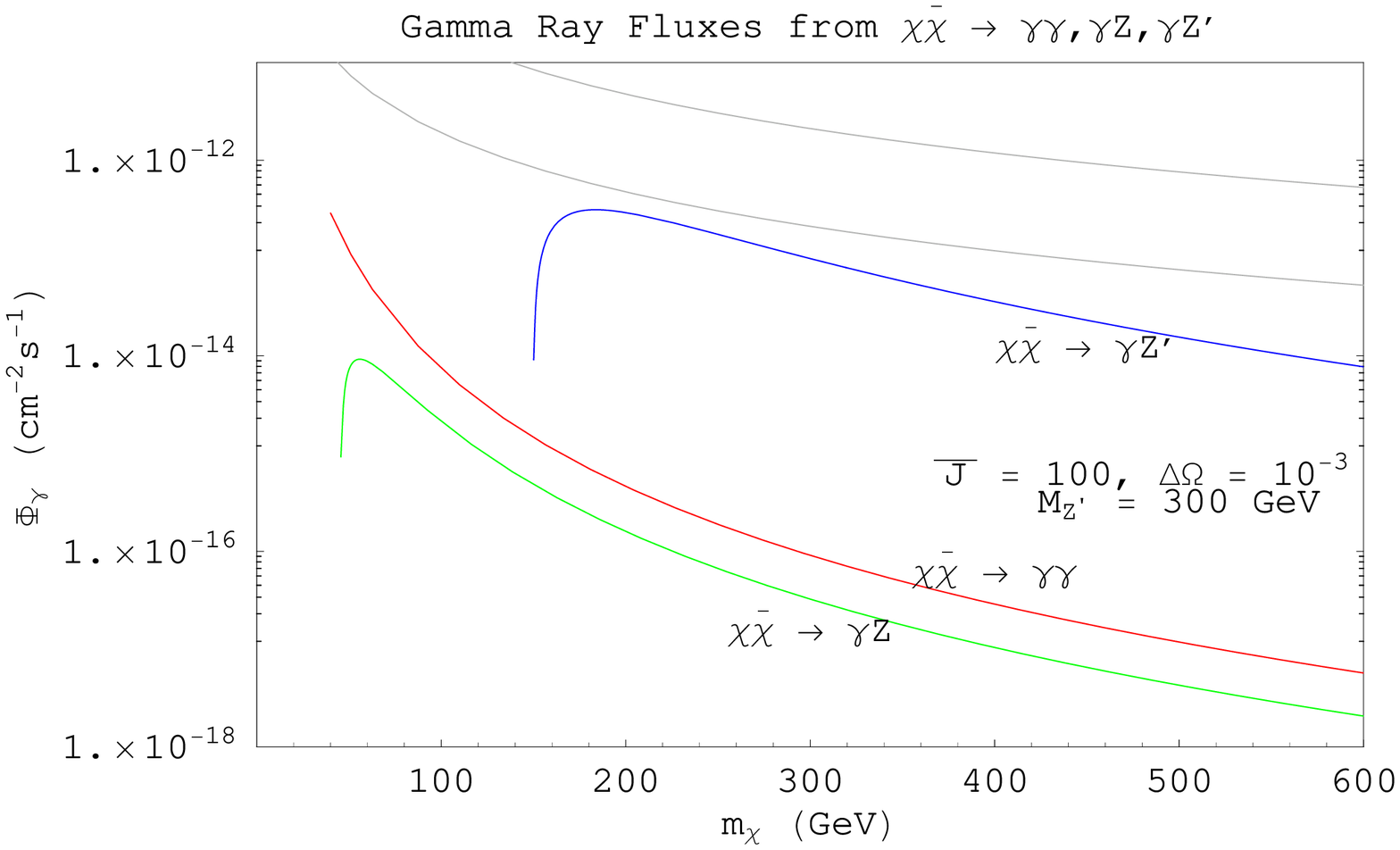}
\caption{Photon flux from $\chi \bar \chi$ annihilation into 
$\gamma \gamma, \gamma Z$ and $\gamma Z'$ near the Galaxy center as a function of $m_\chi$.}
\label{fig:photonflux}    
\end{figure}

After freeze out, pair annihilation of $\chi$ and $\bar \chi$ becomes greatly suppressed.
However, at later times after the structure formation, $\chi \bar \chi$ annihilation 
may regain its importance in the dense region of our universe like the center of the galaxy.
Thus monochromatic photon line from the processes 
$\chi \bar \chi \to \gamma \gamma, \gamma Z, \gamma Z'$ 
could be the ``smoking-gun'' signal of dark matter annihilation at our Galaxy center,  
which may be detected by various ACT array or space satellite experiments.
The gamma-ray flux is given by \cite{hooper} 
\begin{eqnarray}
\Phi_\gamma (\Delta \Omega, E) & \approx &5.6 \times 10^{-12} 
\frac{d N_\gamma}{d E}
\left( \frac{\sigma v} {{\rm pb}} \right)
\left( \frac{1 \, \rm TeV} {m_\chi} \right)^2 \nonumber \\
&& \times {\overline J} (\Delta \Omega) \Delta\Omega \, {\rm cm^{-2} \, s ^{-1}} 
\end{eqnarray}
with $d N_\gamma / d E$ being the photon energy spectrum 
( a Dirac delta function $\delta(E - E_\gamma)$ in our case) and 
the quantity $J(\psi)$ defined by 
\begin{eqnarray}
J(\psi) & = &\frac{1}{8.5\, {\rm kpc}}
\left( \frac{1}{0.3 \, {\rm GeV/cm^3}} \right)^2 \nonumber \\
&& \times \int_{\rm line \; of \; sight} ds \rho^2(r(s,\psi)) \; .
\end{eqnarray}
Note that $J(\psi)$ depends on the halo profile $\rho$ of the dark matter 
and its values may vary over several order of magnitudes \cite{hooper}.
TeV gamma-rays from Sgr A* (a hypothetical super-massive black hole)
near the Galactic center had been observed recently by 
CANGAROO \cite{cangaroo},  HESS \cite{hess} and VERITAS \cite{veritas}.
While these high energy gamma-rays are generally believed to have a violent 
astrophysical origin, they may play the role of continuum background for 
dark matter detection.
Detectability of photon line above continuum background at GLAST \cite{glast} and HESS \cite{hess2} 
was studied in \cite{Zaharijas-Hooper}, where the following criteria was derived
by extrapolating the TeV data to lower energy
\begin{eqnarray}
\label{detectability}
{\rm Photon \; flux} & \ge &
1.9 \times ({\rm TeV}/m_\chi)^2 \nonumber \\
&&  \times ( 10^{-14} - 10^{-13} )\;{\rm cm}^{-2}\;{\rm s}^{-1} \; .
\end{eqnarray}
In Fig.[\ref{fig:photonflux}], the photon fluxes from the aforesaid processes are plotted as a
function of  $m_\chi$ with $m_{Z'} = 300$ GeV. 
One can see that for $m_\chi < 300$ GeV,
the flux from the channel $\chi \bar \chi \to Z' \gamma$ can be quite close to the detectability range 
of (\ref{detectability}) as indicated by the two upper curves of lighted-color.
Note that a rather conservative value of $\bar J = 100$ was used in this plot.

\section{Collider Implication}
\label{collider}

\subsection{Large Hadron Collider (LHC)}
Masses for the Stueckelberg $Z'$ and the hidden fermion $\chi$  
in the range of a few hundred GeV are seen to be compatible with the WMAP data.
Discoveries of these new particles at the LHC or at a future $e^- e^+$ collider are thus feasible.
\begin{figure}
\includegraphics[width=0.5\textwidth,height=0.35\textwidth,angle=0]{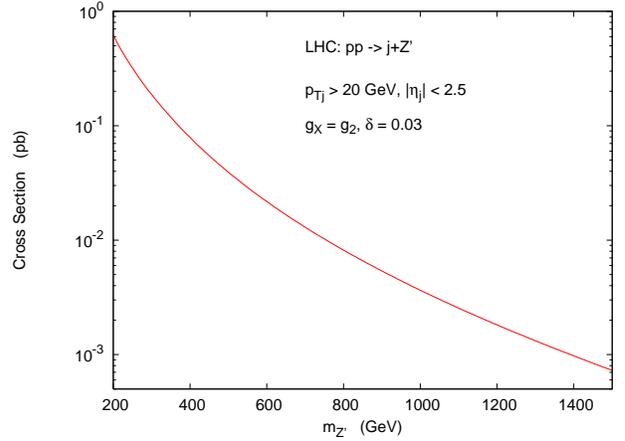}
\caption{Cross section of $p p \to Z'$ + monojet $\to \chi \bar \chi$ + monojet at the LHC.
Branching ratio for $Z' \to \chi \bar \chi$ has been set to be unity.}
\label{fig:lhc}     
\end{figure}
In Fig.[\ref{fig:lhc}], we plot the cross section for singly production of $Z'$ along with a monojet 
due to the parton process $q \bar q \to Z' g$. Since $Z' \to \chi \bar \chi$, the signal would be monojet plus missing energy. The transverse momentum 
$p_{T_j} > 20$ GeV and rapidity $\vert \eta_j \vert < 2.5$ cuts were imposed on the jet. Moreover, the 
input values of $g_X Q_X^\chi$ and $\delta$ are chosen to be those suggested by the above relic density calculation.

\subsection{International Linear Collider (ILC)}
\begin{figure}
\includegraphics[width=0.5\textwidth,height=0.35\textwidth,angle=0]{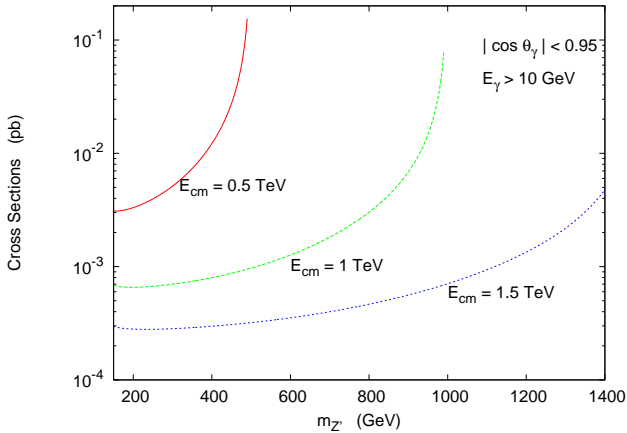}
\caption{Cross section of $e^- e^+ \to Z' \gamma \to \chi \bar \chi + \gamma$ at the ILC of various 
center of mass energies. Branching ratio for $Z' \to \chi \bar \chi$ has been set to be unity.}
\label{fig:ilc}  
\end{figure}

At the ILC, one can look for singly $Z'$ production together with a monochromatic photon with energy fixed by the $m_{Z'}$ mass and center of mass energy. The total cross section as a function of $m_{Z'}$ is plotted in Fig.[\ref{fig:ilc}] 
for three different ILC center of mass energies of 0.5, 1 and 1.5 TeV.
We have imposed the cuts of $E_\gamma > 10$ GeV on the photon energy 
and  $-1 < \cos \theta_\gamma < +1$ for the angle of the photon momentum with 
respect to the beam axis in these curves. As in the previous subsection, 
we chose $g_X Q_X^\chi = g_2$ and $\delta = 0.03$.

\section{Conclusion}
\label{concl}

Adding a pair of Dirac fermion-antifermion in the hidden Stueckelberg $U_X(1)$ sector provides interesting phenomenology for both collider physics and astrophysics. We have demonstrated that if the gauge coupling of the hidden $U_X(1)$ sector has a comparable size like the electroweak gauge group, the hidden fermion carries milli-charge and can be served as an alternative dark matter with a viable relic density suggested by the WMAP data. Pair annihilation of the hidden fermions at the Galactic center can give rise to a ``smoking-gun" signal of monochromatic line that may be above the continuum background and therefore can be probed by the next generation of gamma-rays experiments. 
Moreover, both the hidden fermion and the Stueckelberg $Z'$ can have mass range in a few hundred GeV without any conflict with existing data. If  the decay of $Z'$ into hidden fermion-antifermion pair
is kinematically allowed, this provides a new invisible decay mode for $Z'$ other than the neutrinos. The $Z'$ width is no longer narrow but its dominant decay mode is the invisible mode $\chi \bar \chi$.
Thus it provides new challenge to search for such new neutral gauge boson at the LHC or ILC. 
To search for these new particles,  we propose to use the signals of monojet plus missing energy at the LHC and single photon plus missing energy at the ILC.

\vskip 0.25in 
This work was partially supported by NSC of Taiwan.

\end{document}